\documentclass[trackchanges]{aastex7}

\begin{document}

\title{Expanding the \texttt{SPISEA} Stellar Population Synthesis Software to the Substellar Regime}

\author[orcid=0009-0006-8866-4224,sname='Begbie']{Caitlin Begbie}
\affiliation{Department of Astronomy, University of California,
Berkeley, 501 Campbell Hall, Berkeley, CA 94720, USA}
\email[show]{caitlinbegbie@berkeley.edu}  

\author[orcid=0000-0001-6279-0552]{Rachel Street}
\affiliation{Las Cumbres Observatory, 6740 Cortona Drive, Suite 102, Goleta, CA 93117, US}
\email{rstreet@lco.global}

\author[orcid=0000-0002-2729-5369]{Katarzyna Kruszyńska}
\affiliation{Las Cumbres Observatory, 6740 Cortona Drive, Suite 102, Goleta, CA 93117, US}
\email{kkruszynska@lco.global}

\author[orcid=0000-0001-9611-0009]{Jessica Lu}
\affiliation{Department of Astronomy, University of California,
Berkeley, 501 Campbell Hall, Berkeley, CA 94720, USA}
\email{jlu.astro@berkeley.edu}

\author[orcid=0000-0003-2874-1196]{Matthew Hosek Jr.}
\affiliation{Physics and Astronomy Department, University of California, Los Angeles, CA 90095-1547, USA}
\email{mwhosek@gmail.com}

\author[orcid=0000-0002-0287-3783]{Natasha Abrams}
\affiliation{Department of Astronomy, University of California,
Berkeley, 501 Campbell Hall, Berkeley, CA 94720, USA}
\email{nsabrams@berkeley.edu}

\author[orcid=0000-0003-4591-3201]{Macy Huston}
\affiliation{Department of Astronomy, University of California,
Berkeley, 501 Campbell Hall, Berkeley, CA 94720, USA}
\email{mhuston@berkeley.edu}

\begin{abstract}

We present an extension of the \texttt{SPISEA} stellar population synthesis framework that adds brown dwarfs to the existing range of stellar mass objects, enabling physically consistent modeling of brown dwarfs within synthetic star clusters. Previous versions of \texttt{SPISEA} included limited substellar support, relying on outdated initial mass functions and incomplete atmospheric and evolutionary coverage below the hydrogen-burning limit. This was addressed through the implementation of a modern substellar initial mass function based on robust observational constraints, the introduction of merged atmospheric grids that smoothly transition between stellar and brown dwarf regimes, and the construction of  unified evolutionary tracks spanning the lowest-mass brown dwarf objects through massive stars at solar metallicity. The updated framework was validated by comparing simulated color–magnitude diagrams to observational data from the Pleiades, Upper Scorpius, and M44 clusters using Gaia, UKIDSS, and 2MASS photometry. The new models allow for generation of user-specified isochrones and clusters that reproduce observed stellar behaviors while enabling realistic population synthesis in the brown dwarf regime. This work extends \texttt{SPISEA}’s applicability to substellar science cases, including young cluster studies and microlensing simulations, and provides a foundation for future incorporation of planetary-mass objects and non-solar metallicities.

\end{abstract}



\section{Introduction} 

Stellar clusters are essential testbeds for theories of star formation and evolution because they are a self-consistent, co-evolutionary population with measurable properties such as age, metallicity, and extinction \citep{fernandes2010testing}. Modeling stellar populations in galaxies is a key tactic to both gain insight into the limits of our knowledge, such as the process of galaxy formation as a function of cosmic time, and decipher real cluster observations \citep{2013IAUS..295..272M}.


The majority of stellar clusters can be modeled as simple stellar clusters (SSPs), which are characterized by a single age and metallicity \citep{maraston2003stellar}. As such, recent years have introduced various packages to model unresolved SSPs, including \texttt{Starburst99} \citep{leitherer1999starburst99}, \texttt{Flexible Stellar Population Synthesis} (\texttt{FSPS}; \citealt{conroy2009propagation, conroy2010propagation}), and \texttt{Stoichiastically Lighting Up Galaxies} (\texttt{SLUG}). The \texttt{Stellar Population Interface for
Stellar Evolution and Atmospheres} (\texttt{SPISEA}; \citealt{hosek2020spisea})\footnote{https://github.com/astropy/SPISEA} is another such open-source Python package to generate SSPs, but allows for increased user manipulation through a wide range of configurable parameters, as well as options to generate both unresolved and resolved clusters. In this case, resolved clusters allow for the assignment of individual object properties, and unresolved clusters are represented as a composite spectrum. With \texttt{SPISEA}, users have control over numerous cluster parameters, including age, metallicity, distance, extinction, differential extinction, cluster mass, initial mass function (IMF), extinction law, multiplicity properties, evolutionary model, atmospheric model, photometric filters, and initial-final-mass-relation (IFMR); along with the ability to generate compact objects and binary populations \citep{hosek2020spisea}. The pipeline for \texttt{SPISEA}'s isochrone and cluster object generation is shown in Figure \ref{fig:1}, and outlined in \citet{hosek2020spisea}. 

Along with overall cluster statistics, simulating stellar populations heavily relies on individual object generation and properties. The original focus of \texttt{SPISEA} was to statistically produce stellar objects from the M dwarf range to the most massive G stars ($\sim 0.1 - 120 M_\odot$), with associated atmospheric and evolutionary characteristics, as both primary and companion objects in singular, binary, and higher-order systems. Extending this lower limit into the substellar regime (where brown dwarfs and planets are found) was omitted in previous releases. These low-mass objects are defined by complex, cool, and high-density environments, which require accounting for molecular chemistry, cloud formation, and non-equilibrium processes, which makes their physical processes and simulation mechanisms distinct from stellar objects \citep{allard2012models}.

Brown dwarfs in particular occupy a critical regime at the boundary between stars and planets, bridging the gap between hydrogen-burning objects and giant planets ($\sim0.01 - 0.08 M_\odot$). Current estimates suggest that they are potentially as numerous as stars, meaning that their formation, frequency, and evolution encode valuable information about star formation processes, cluster demographics, and the overall composition of the universe \citep{zuckerman2000brown}. Over the past two decades, large-area infrared surveys such as 2MASS \citep{skrutskie2006two}, WISE \citep{wright2010wide}, UKIDSS \citep{lawrence2007ukirt}, and Pan-STARRS \citep{chambers2016pan} have revealed substantial new populations of brown dwarfs in the field and in nearby clusters \citep{kirkpatrick1999dwarfs, scholz2012ukidss, zhang2018pan}, motivating increasingly detailed theoretical models of substellar atmospheres and evolution \citep{phillips2020new}. 

Accurate modeling of the brown dwarf population within the context of stellar clusters would enable new studies of environments where substellar objects contribute significantly to the observed population, such as young clusters, deep infrared surveys, and microlensing simulations \citep{bejar2018brown}. However, despite recent observational advances, modeling brown dwarfs within stellar population synthesis frameworks remains challenging. The substellar regime is characterized by complex atmospheric chemistry, dust formation, and partially degenerate interiors that evolve primarily through cooling rather than sustained nuclear burning -- all of which complicate the construction of continuous mass–luminosity and color–magnitude relations \citep{phillips2020new}. As a result, SSP modeling packages, including aforementioned \texttt{Starburst99} \citep{leitherer1999starburst99}, \texttt{FSPS} \citep{conroy2009propagation, conroy2010propagation}, and \texttt{SLUG} \citep{da2012slug, krumholz2015slug} have historically only focused on stellar populations, treating brown dwarfs in an overly simplified or inconsistent manner, or opting to exclude them altogether.

In this work, we extended \texttt{SPISEA} to enable physically consistent population synthesis from high-mass stars through the brown dwarf range, making it the first SSP capable of doing so. We describe the software upgrades necessary to include brown dwarfs in simulation procedures, which expands \texttt{SPISEA}'s applicability to substellar populations without compromising the original capabilities of the package.
The resulting framework enables direct comparison between simulated clusters and observations from optical surveys, such as Gaia \citep{prusti2016gaia}, and infrared surveys sensitive to brown dwarfs, including WISE \citep{wright2010wide}, UKIDSS \citep{lawrence2007ukirt}, 2MASS \citep{skrutskie2006two}, JWST \citep{chen2025brown} and, in the future, Roman \citep{wang2022high}.


This paper is organized as follows. In Section \ref{sec:2} we outline the \texttt{SPISEA} package, noting the elements of interest that were changed. In Section \ref{sec:3}, we present comparisons between simulated clusters and observational data, and in Section \ref{sec:4}, we summarize our conclusions and outline future extensions.

\section{\texttt{SPISEA} Overview and Updates}
\label{sec:2}

\subsection{\texttt{SPISEA} Overview}
\label{subsec:2.1}

\begin{figure}
    \centering
    \includegraphics[width=\linewidth]{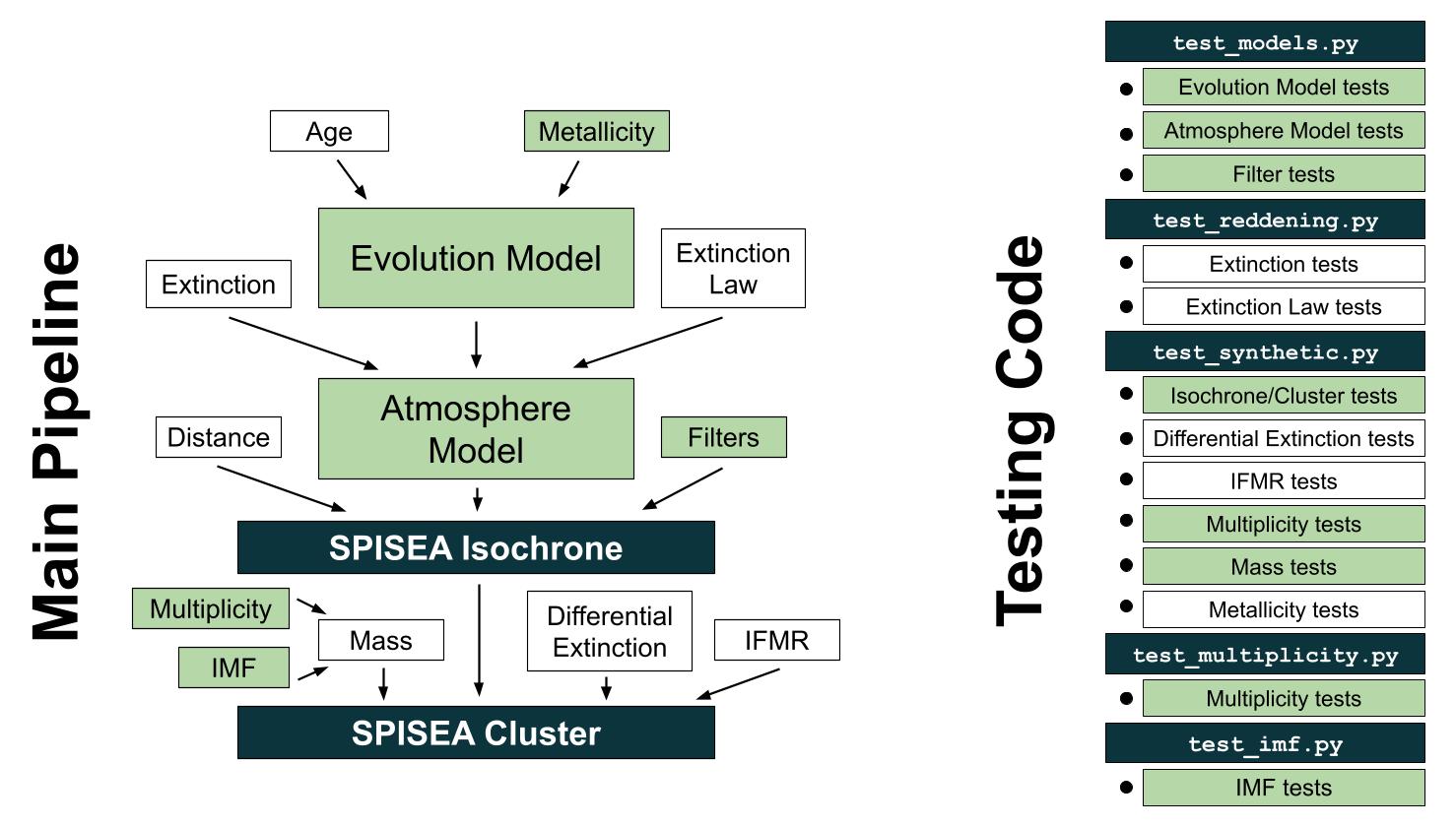}
    \caption{Diagram of the main pipeline of the \texttt{SPISEA} code (left), as well as associated testing with each feature (right). In the main pipeline, the white and light green boxes represent inputs specified by the user, where light green also represents the features updated in this current work. The primary outputs of the code, the isochrone and the cluster, are shown in the dark boxes. In the testing code, the dark boxes show the testing code files included in \texttt{SPISEA} that are related to the user-specified features. The light green boxes represent testing that was changed to reflect and validate the updates in this work. This figure is adapted from \citet{hosek2020spisea}.}
    \label{fig:1}
\end{figure}

SPISEA is an open-source Python package for generating SSPs, characterized by its modular design that allows user control over 13 distinct characteristics: age, metallicity, evolution model, extinction, extinction law, atmosphere model, distance, filters, multiplicity, IMF, mass, differential extinction, and the IFMR. Figure \ref{fig:1} shows a diagram of the components of the software and testing suite, illustrating those upgraded during this work.
For this paper, we will focus on the features that were updated, that is metallicity considerations, evolution and atmosphere models, multiplicity constraints, and the IMF. Information about the other features is available in \citet{hosek2020spisea}.

\subsection{Metallicity}
\label{subsec:2.2}
One of the main functions of \texttt{SPISEA} is to create a SSP at a specified age, distance, extinction and metallicity. Metallicity is an important factor in stellar clusters, as objects with different heavy element concentrations follow distinct evolutionary and atmospheric pipelines \citep{gehrig2023influence}. \texttt{SPISEA} assigns intrinsic properties to individual stars through metallicity-dependent stellar atmospheric and evolutionary grids, which span a wide range of values. For our purposes in this work however, updates were solely made for isochrones and clusters of solar metallicity values ($Z=0.0$) due to the lack of availability of substellar models at other metallicities. This means that, at this time, brown dwarfs are not generated for non-solar metallicities.

\subsection{Initial Mass Function}
\label{subsec:2.3}
An IMF statistically describes the number of objects expected to be formed in each bin of a stellar mass distribution and is typically expressed as a power law:

\begin{equation}
    \frac{dN}{dM} \propto M^{-\alpha}
    \label{eq:powerlaw}
\end{equation}

\noindent where $N$ is the number of objects, $M$ is the mass of the objects, and $\alpha$ is the slope parameter of the mass function. Since its introduction in \cite{salpeter1955luminosity}, the IMF has been expanded into a “broken power law” format in hopes of addressing slope changes across more expansive stellar mass regimes \citep{kroupa2001variation}. \texttt{SPISEA} supports both a user-specified IMF, where the user can list mass breaks and their associated slope parameters, as well as several pre-determined IMFs, such as from Salpeter \citep{salpeter1955luminosity}, Miller \& Scalo \citep{miller1979initial}, Kennicut \citep{kennicutt1983rate}, Kroupa \citep{kroupa2001variation}, and Weidner \& Kroupa \citep{weidner2004evidence}.

\begin{figure}
    \centering
    \includegraphics[width=0.9\linewidth]{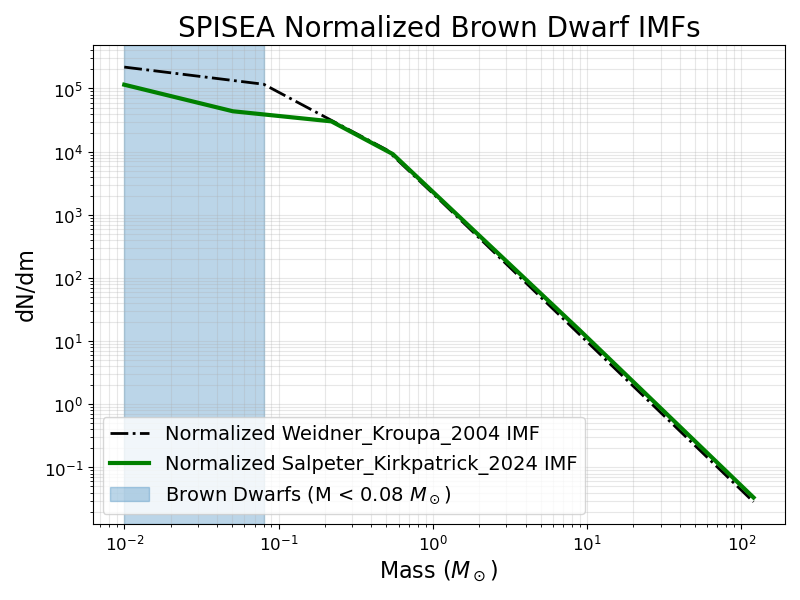}
    \caption{Normalized broken power-law initial mass functions (IMFs) in \texttt{SPISEA} defining object generation from 0.01 - 120 M$_\odot$ as per Equation \ref{eq:powerlaw}. \texttt{Weidner\_Kroupa\_2004} represents the pre-existing substellar model in \texttt{SPISEA}, defining the entire brown dwarf regime with $\alpha$ = 0.3, as per \citet{weidner2004evidence}. \texttt{Salpeter\_Kirkpatrick\_2024} is the newly added substellar IMF, with updated $\alpha$ values of 0.6 and 0.25 for brown dwarf mass ranges below and above 0.05 M$_\odot$, respectively. These indices more accurately reflect the observed flattening of mass generation in the low-mass star/high-mass brown dwarf regime before increasing again in the low-mass brown dwarf regime, as outlined in \cite{kirkpatrick2024initial}.}
    \label{fig:imf}
\end{figure}

Previous iterations of the \texttt{SPISEA} codebase incorporated an IMF extending into the brown dwarf mass range, based on \cite{weidner2004evidence}. However, the slope cited for the brown dwarf regime ($\alpha = +0.3$ for $0.01 M_\odot < M < 0.08 M_\odot$) is considered outdated as more recent studies targeting substellar objects in young clusters and fields identify a steeper slope, namely $\alpha \gtrsim +0.5$ \citep{best2024volume, muvzic2019looking, kirkpatrick2021field, kirkpatrick2024initial}. Thus, modernizing \texttt{SPISEA} to generate brown dwarfs in accurate proportions involved the implementation of a new IMF class, \texttt{imf.Salpeter\_Kirkpatrick\_2024}, which utilizes updated slope values from \cite{kirkpatrick2024initial} to describe generated masses from $0.01 M_\odot$ to $8.0 M_\odot$. This work was chosen as our baseline due to its unique IMF construction from a sample of about 3600 stars and brown dwarfs – allowing empirical proof in constraining slope values down to 0.01 $M_\odot$. From 8.0 $M_\odot$ to 120 $M_\odot$, the IMF retains the canonically recognized slope of $\alpha = +2.35$, originally posed in \cite{salpeter1955luminosity} and observationally reinforced by numerous authors since, such as \cite{2014ARA&A..52..415M}. A strict upper mass limit of 120 $M_\odot$ was imposed based on current theoretical stellar structure limits and a lack of observational evidence addressing bodies past this size, suggesting that more massive stars are either incredibly rare or potentially non-existent in resolved clusters \citep{figer2005upper, crowther2010r136}. Between 0.01 to 0.05 $M_\odot$, the slope is recognized as $\alpha=0.6$, and above this range $\alpha=0.25$. This suggests that the rate of production of brown dwarfs diminishes between low mass stars to high mass brown dwarfs, before increasing again in the low mass brown dwarf regime. Table \ref{tab:1} illustrates the different stellar mass bins and slope parameters used for the \texttt{Salpeter\_Kirkpatrick\_2024} IMF, and Figure \ref{fig:imf} shows the new, normalized \texttt{Salpeter\_Kirkpatrick\_2024} IMF against the previously included substellar IMF, \texttt{Weidner\_Kroupa\_2004}, which reflects a steeper substellar power-law index and the observed turnover at $\sim 0.05 M_\odot$. 

\begin{table}[htbp]
\centering
\begin{tabular}{|c|c|}
\hline
\textbf{Mass Range (M$_\odot$)} & \textbf{Slope Parameter ($\alpha$}) \\ \hline
0.01 - 0.05 & 0.6                \\ \hline
0.05 - 0.22  & 0.25                \\ \hline
0.22 - 0.55  & 1.3                \\ \hline
0.55 - 8.0  & 2.3                \\ \hline
8.0 - 120  & 2.35                \\ \hline
\end{tabular}
\caption{The newly implemented \texttt{Salpeter\_Kirkpatrick\_2024} IMF extending to the substellar regime as described in Section \ref{subsec:2.3}, illustrating the different stellar mass bins and associated slope parameters ($\alpha$) as per Equation \ref{eq:powerlaw}.}
\label{tab:1}
\end{table}

\subsection{Evolution Models}
\label{subsec:2.5}
Stellar properties are highly dependent on their phase of evolution. Fundamental properties (e.g. temperature, luminosity, and surface gravity) of a star change throughout its lifetime as a result of the chemical processes happening within \citep{aguirre2017stellar}. While larger stars eventually evolve into compact remnants (e.g. white dwarfs, neutron stars, and black holes), because brown dwarfs do not sustain hydrogen fusion, their evolution is defined by monotonic cooling at a nearly constant mass \citep{ryan2022self}. 

\texttt{SPISEA} handles evolutionary properties through a network of evolution model grids, which the user specifies upon isochrone generation. The previous iteration of \texttt{SPISEA} introduced several options of evolutionary model families, including MESA Isochrones and Stellar Tracks (MIST; \citealt{choi2016mesa}), Geneva \citep{ekstrom2012grids}, and PAdova \& TRieste Stellar Evolution Code (PARSEC; \citealt{bressan2012parsec}), as well as hybrid grids that allow for the combination of models across parameter space to account for the individual strengths of each. One such hybrid grid, \texttt{MergedBaraffePisaEkstromParsec}, makes use of four well-supported evolutionary models for solar metallicities: Baraffe15 \citep{baraffe2015new}, which covers low mass pre-main sequence stars; Pisa \citep{tognelli2011pisa}, for intermediate-mass pre-main-sequence and main-sequence stars; and the Ekstrom/Geneva (both with and without rotation; \citealt{ekstrom2012grids}) and Parsec v1.2 \citep{bressan2012parsec} models for higher-mass and older main-sequence evolution.

To maintain continuity and ensure compatibility between atmospheric and evolutionary models, for the substellar regime, we applied solar-metallicity evolutionary models from \citet{phillips2020new}, computed using the same ATMO framework as our newly-implemented atmospheric model grids (Section \ref{subsec:2.4}). These models describe how brown dwarf temperatures, luminosities, and surface temperatures change as a function of their initial masses, over time. We also applied the canonical assumption that brown dwarf mass loss is negligible over their lengthy lifespans, such that their present-day masses are equivalent to their initial masses \citep{burrows1993science, cirkovic2005brown}. Taking these models and assumptions into account, we created a new unified evolutionary grid, \texttt{MergedPhillipsBaraffePisaEkstromParsec}, spanning masses from 0.01 M$_\odot$ to the upper end of the stellar mass function ($\sim$120 M$_\odot$), and ages from log(age/year) = 6.0 - 10.1 at solar metallicity. For young ages (logAge $\leq$ 8.0), the pre-main sequence regime is constructed from three models: Phillips (0.01 - 0.07 M$_\odot$), Baraffe (0.075 - 0.4 M$_\odot$), and Pisa ($\geq$ 0.5 M$_\odot$). Between the models, weighted interpolation was performed in the overlapping regions (0.07 - 0.075 M$_\odot$ for Phillips to Baraffe, and 0.4 - 0.5 M$_\odot$ for Baraffe to Pisa) to allow for smooth continuity across model transitions. For the main sequence, Ekstrom models are adopted for log ages $\leq$ 7.4, while PARSEC models are used at later times. For log age $\textgreater$ 8, the evolution is dominated by the Phillips and PARSEC grids. Table \ref{tab:3} documents the properties of the new \texttt{MergedPhillipsBaraffePisaEkstromParsec} model grid.

\begin{table}[htbp]
\textbf{\texttt{MergedPhillipsBaraffePisaEkstromParsec}: log(Age) $<$ 7.4}
\centering
\begin{tabular}{|c|c|c|}
\hline
\textbf{Mass Range ($M_\odot$)} & \textbf{Evolution Model} & \textbf{Reference} \\ \hline
\textit{0.01 - 0.070} & \textit{Phillips} & \textit{\citet{phillips2020new}}  
\\ \hline
\textit{0.070 - 0.075}  & \textit{Phillips/Baraffe transition}  & \textit{\citet{phillips2020new, baraffe2015new}}           
\\ \hline
0.075 - 0.4  & Baraffe & \citet{baraffe2015new} 
\\ \hline
0.4 - 0.5  & Baraffe/Pisa transition & \citet{baraffe2015new, tognelli2011pisa}
\\ \hline
0.5 - 120  & Pisa \& Geneva & \citet{tognelli2011pisa, ekstrom2012grids} 
\\ \hline
\end{tabular}
\medskip

\textbf{\texttt{MergedPhillipsBaraffePisaEkstromParsec}: log(Age) $>$ 7.4}
\centering
\begin{tabular}{|c|c|c|}
\hline
\textbf{Mass Range ($M_\odot$)} & \textbf{Evolution Model} & \textbf{Reference} \\ \hline
\textit{0.01 - 0.075} & \textit{Phillips} & \textit{\citet{phillips2020new}}  
\\ \hline
\textit{0.075 - 0.2}  & \textit{Phillips/PARSEC transition} & \textit{Interpolation, Section \ref{subsec:2.5}}     
\\ \hline
0.2 - 120  & Geneva/PARSEC & \citet{ekstrom2012grids, bressan2012parsec} 
\\ \hline

\end{tabular}
\caption{The new \texttt{MergedPhillipsBaraffePisaEkstromParsec} model grid, now including substellar considerations. There are two age regimes, pre-main sequence (log(Age) $<$ 7.4; top) and main-sequence (log(Age) $>$ 7.4; bottom). In each case, evolutionary model grids are assigned to objects based on their masses, which are used to provide stars with $T_{\rm eff}$ and surface gravity values. In both tables, the new additions are in italics.}
\label{tab:3}
\end{table}

It is worth noting that, due to different model grids addressing distinct mass regimes, there was a gap present between the Phillips model and the Pisa and PARSEC models, spanning a range of $0.075 - 0.2 M_\odot$. Thus, to ensure a physically smooth transition across the stellar-substellar boundary and pre-main sequence to main sequence regimes, we adopted regression-based methods to interpolate across this mass gap for all ages. In the Phillips-Pisa transition region, Gaussian Process regression with \texttt{scikit learn}'s \citep{scikit-learn} Matern kernel was used to model the mass vs. luminosity, effective temperature, and logarithmic surface gravity relations at fixed age intervals, allowing for a continuous sequence with smooth first derivatives. To determine the optimal parameters for the Gaussian Process interpolation, we performed a grid search over the Matern kernel parameter space, varying both the characteristic length scale ($\ell$) and smoothness parameter ($\nu$). For each pair of parameters, we evaluated model performance using 5-fold cross-validation. The dataset was partitioned into five subsets; in each iteration, four subsets were used for training and the remaining subset for validation. Predictions were generated for the validation set, and residuals were computed as the squared differences between predicted ($y_{i, \rm pred}$) and true ($y_{i}$) values. We then defined the total chi-squared ($\chi^2$) value to be given by 

$$\chi^2 = \sum_{i=1}^{N}(y_i-y_{i, \text{pred}})^2$$

\noindent assuming uniform uncertainties across all data points. To then account for model complexity, we took the reduced chi-squared statistic,

$$\tilde{\chi}^2 = \frac{\chi^2}{N-k}$$

\noindent where $N$ is the total number of points used for validation, and $k=2$ represents the number of kernel parameters. We then repeated this procedure across the full parameter grid, and selected the optimal model by minimizing $\log_{10}(\tilde{\chi}^2)$. 

\begin{figure}
    \centering
    \includegraphics[width=0.9\linewidth]{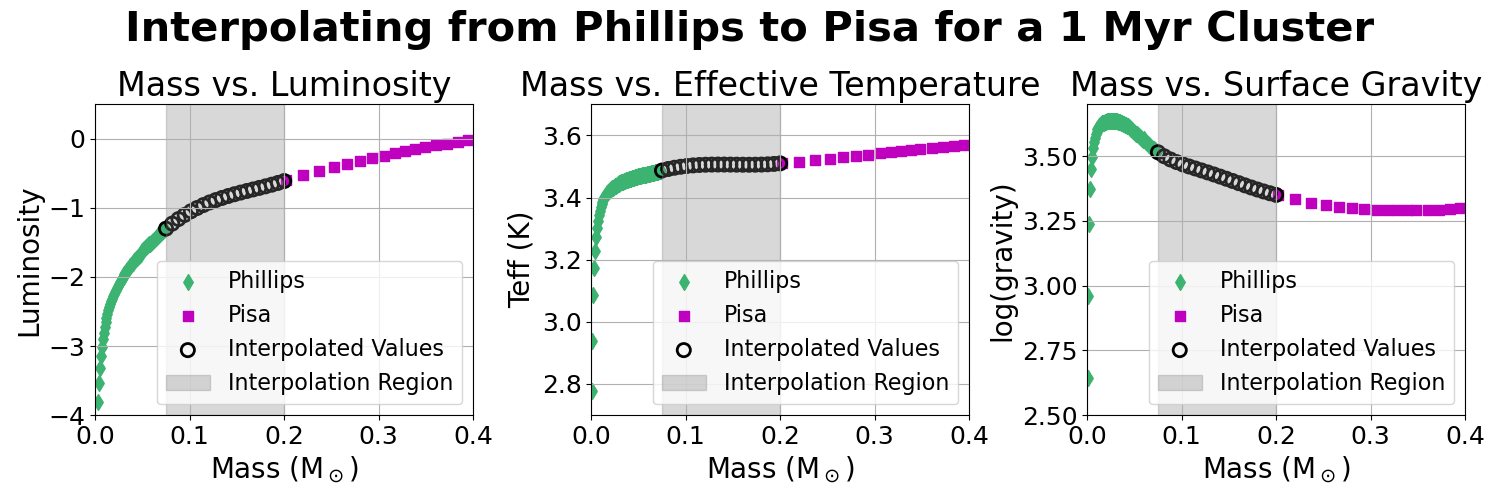}
    \includegraphics[width=0.9\linewidth]{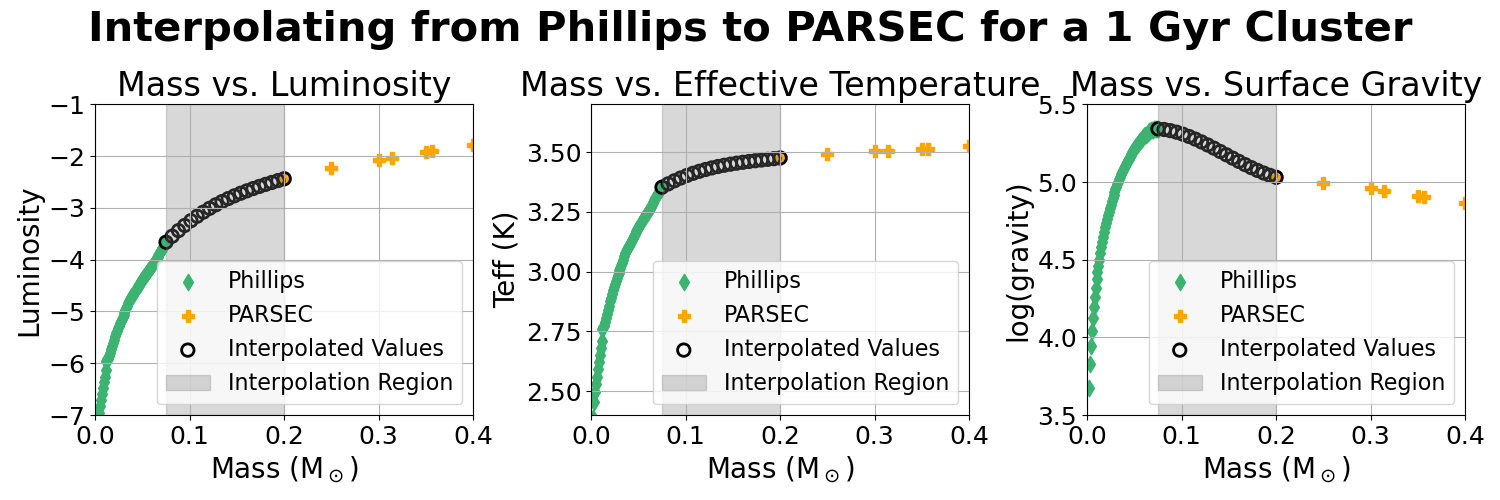}
    \caption{Example results of interpolated evolutionary models over mass gap regions between the \cite{phillips2020new} models and the pre-existing Pisa \citep{tognelli2011pisa} and PARSEC \citep{bressan2012parsec} models. Gaussian Process regression methods and Hermite splines were employed to smoothly define values in the intermediary region (0.075 - 0.2 M$_\odot$) between models, which the literature does not cover. The values in the interpolated regions are flagged in the final star systems tables, to inform users that their values are not yet observationally confirmed.}
    \label{fig:4}
\end{figure}

For the Phillips-PARSEC merger at older ages, the mass vs. luminosity and temperature relations were more monotonic and well-behaved. This allowed for the use of Hermite splines to preserve shapes, which was less computationally expensive. Due to more complex deviations between models for surface gravity relations, Gaussian Process regression was retained for thorough coverage. This allowed for the omission of synthetic features not present in literature, such as increased curvature in the mass gap region, that emerged as a side-effect of fitting procedures. Examples of the Phillips-Pisa and Phillips-PARSEC mergers, can be seen in Figure \ref{fig:4} for a 1 Myr- and 1 Gyr-aged cluster, respectively.

All model parameter values derived from interpolation are explicitly flagged in the output isochrone tables, allowing users to distinguish original model values from regression-based transition regions. This is important because, while the interpolated regime represents the best current theoretical estimate of evolutionary behaviors in this understudied region, it will require readdressing when further models become available. Overall, however, the final \texttt{MergedPhillipsBaraffePisaEkstromParsec} isochrones provide continuous, solar-metallicity evolutionary sequences in mass, luminosity, effective temperature, and surface gravity from planetary-mass objects through massive stars, enabling the generation of synthetic clusters with brown dwarfs and low-mass stars without artificial discontinuities at the hydrogen burning limit.

\subsection{Atmosphere Models}
\label{subsec:2.4}
After generating substellar masses and assigning fundamental parameters, it was necessary to assign physically consistent atmospheric properties to brown dwarfs. Atmospheric properties are typically determined through incorporation of atmospheric model grids, which attempt to characterize the composition and evolution of object atmospheres based on their temperatures, surface gravities, and metallicities. Substellar atmosphere models are typically derived through detailed comparisons of observed spectra to grids of self-consistent forward models, or through MCMC-driven retrievals of these spectra \citep{karalidi2021sonora}. Because atmospheric modeling in the substellar regime remains an active area of development, the range of available and well-validated model grids remains limited, particularly for non-solar metallicities. 

\texttt{SPISEA} assigns atmospheric parameters through families of atmospheric model grids, which selects values based on the object's effective temperature, surface gravity, and metallicity. These object values are derived from user-specified cluster values, the initial mass of each simulated object, and the object's position along the associated evolutionary track (discussed in Section \ref{subsec:2.5}). The model grids are combined into a single function in \texttt{SPISEA}, \texttt{get\_merged\_atmosphere}, where various atmosphere models built to cover distinct temperature ranges are assigned. Overlapping temperature ranges between model grids are covered by a weighted interpolation between the two models. Previous iterations of \texttt{SPISEA} lacked atmospheric coverage for substellar objects, necessitating the addition of brown dwarf atmosphere models, and expansion of the \texttt{get\_merged\_atmosphere} function into this regime.

For this work we evaluated several reputable substellar model families to assign brown dwarf parameters, such as Sonora \citep{marley2021sonora}, BT-Settl \citep{allard2013bt}, AMES-Cond \citep{allard2000tio, allard2001limiting}, and ATMO 2020 \citep{phillips2020new}. Ultimately, we decided to implement ATMO-based grids developed by \citet{phillips2020new} and \citet{meisner2023exploring} due to their robust range of ages and masses, and precision in modeling comprehensive molecular opacities, vertical mixing, and non-equilibrium chemistry. More specifically, the ATMO models incorporate a number of advances in the treatment of substellar atmospheres compared to earlier grids, including a self-consistent radiative–convective equilibrium framework, updated molecular and atomic opacities, and the inclusion of cloud formation using rainout \citep{phillips2020new}. These improvements allow the ATMO grids to supersede earlier widely used models particularly in reproducing observed spectral features of late-type brown dwarfs. In this initial update, we restrict our grids to the chemical equilibrium versions of the models, as non-equilibrium chemistry and vertical mixing remain active areas of research \citep{karalidi2021sonora}. 

The \cite{phillips2020new} and \cite{meisner2023exploring} grids span effective temperatures from approximately 200 to 3000 K, surface gravities 2.5 $\leq$ log g $\leq$ 5.5, and masses from 0.001 to 0.075 M$_\odot$. Notably, however, the validity of these grids are limited to $T_{\rm eff}$ $\lesssim$ 2000 K, as some high-temperature opacity sources are not included in the ATMO calculations. The \cite{meisner2023exploring} extension specifically targets the coldest and lowest-metallicity brown dwarfs, providing improved coverage of the late-T and Y dwarf regime. Presently, reliable grids at non-solar metallicities remain sparse and are rarely consistent with available evolutionary models, meaning this work solely focuses on solar-metallicity atmospheres.

\begin{figure}
    \centering
    \includegraphics[width=\linewidth]{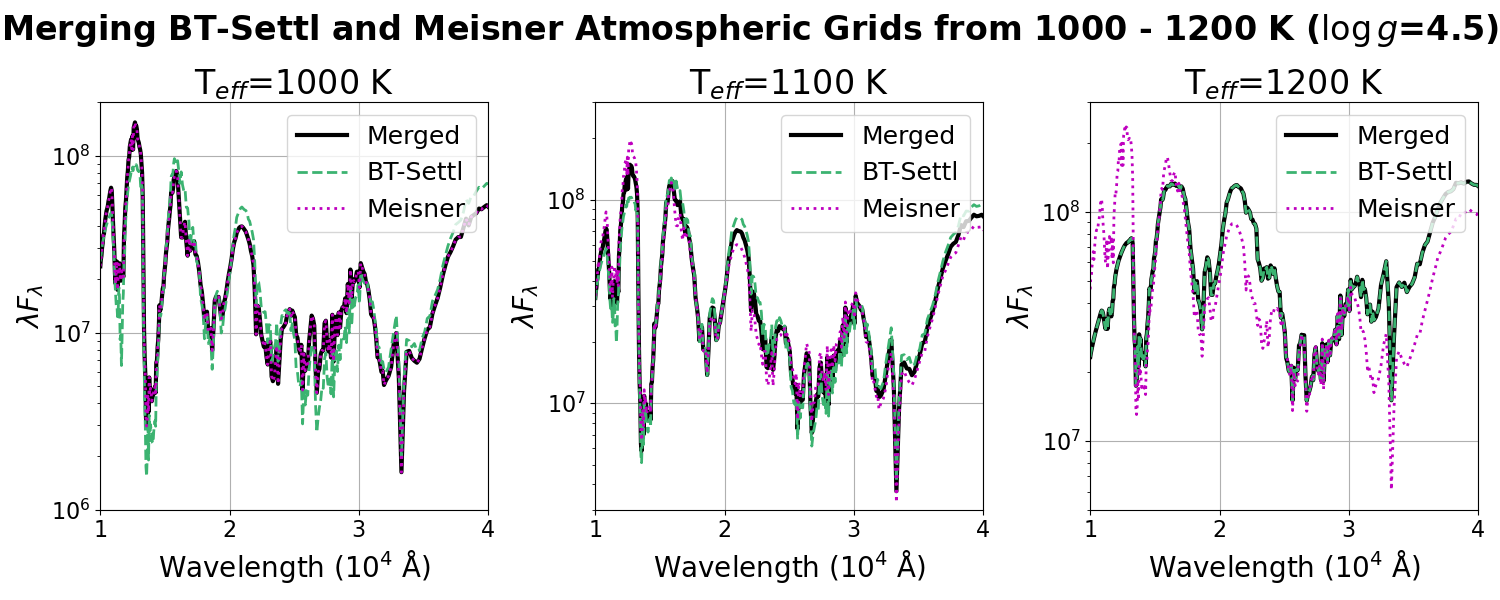}
    \caption{Representation of flux as a function of wavelength for the merged atmosphere model (grey solid) versus BT-Settl (green dashed) and Meisner (magenta dotted) atmospheres for effective temperatures of 1000 K (left), 1100 K (middle), and 1200 K (right). This fully covers the temperature range (1000 - 1200 K) where the preexisting BT-Settl model \citep{allard2012models, baraffe2015new} and newly-added Meisner model \citep{meisner2023exploring} were merged via weighted interpolation. As described in Section \ref{subsec:2.4}, at 1000 K the weighted interpolation model favors the Meisner grid, which is more accurate at lower temperatures. At 1200 K, the model follows the BT-Settl model. At 1100 K, the two models are equally weighted.}
    \label{fig:3}
\end{figure}

To enable a smooth transition between the warmer BT-Settl atmospheres \citep{allard2012models, baraffe2015new} and the cooler ATMO-based Meisner grid within \texttt{SPISEA}, we implemented a new merged atmospheric class, \texttt{merged\_BTSettl\_meisner}. In the temperature range from 1000 - 1200 K where both models provide coverage, spectra were combined via weighted interpolation for all available surface gravity values, as shown in Figure \ref{fig:3} for $\log(g)=4.5$. Specifically, the contribution of the BT-Settl models decreases from unity at 1200 K to zero at 1000 K, with the opposite weighting applied to the Meisner spectra. Outside this transition region, the pre-existing BT-Settl models are used at higher temperatures and the newly-implemented Meisner models are used at lower temperatures. This approach ensures continuity in the assigned spectra and avoids artificial discontinuities across the model transition. The merged spectra and corresponding catalog files were incorporated into \texttt{SPISEA}'s repository, and new atmospheric retrieval routines were implemented and tested for use in synthetic cluster simulations. Table \ref{tab:2} shows the breakdown of the final, expanded \texttt{get\_merged\_atmosphere} function with the new additions.

\begin{table}[htbp]
\centering
\begin{tabular}{|c|c|c|}
\hline
\textbf{$T_{\rm eff}$ Range (K)} & \textbf{Atmospheric Class} & \textbf{Reference} \\ \hline
20000 - 5500 & ATLAS & \citet{castelli2004new}  
\\ \hline
5500 - 5000  & ATLAS/PHOENIXv16 merge & \citet{castelli2004new, husser2013new}           
\\ \hline
5000 - 3800  & PHOENIXv16 & \citet{husser2013new} 
\\ \hline
3800 - 3200  & PHOENIXv16/BTSettl merge & \citet{husser2013new, baraffe2015new}
\\ \hline
3200 - 1200  & BTSettl & \citet{baraffe2015new} 
\\ \hline
\textit{1200 - 1000} & \textit{BTSettl/Meisner2023 merge} & \textit{\citet{baraffe2015new, meisner2023exploring}}
\\ \hline
\textit{1000 - 250} & \textit{Meisner2023} & \textit{\citet{meisner2023exploring}}
\\ \hline
\end{tabular}
\caption{The expanded \texttt{get\_merged\_atmosphere} model grid, now including substellar considerations. Temperature ranges are covered by model grids designed for objects in that regime, with interpolated regions between models. Interpolating between models in overlapping regions ensures continuity over the entire temperature spectrum, and avoids synthetic discontinuities between model grids. The newly implemented Meisner2023 \citep{meisner2023exploring} model, along with the interpolated model between BTSettl \citep{baraffe2015new} and Meisner grids, are shown in italics. Notably, for $T_{\rm eff} \leq 3200$, and solar metallicity, simulations are limited to logg $\geq$ 2.5. Otherwise, the PHOENIXv16 \citep{husser2013new} is applied for all $T_{\rm eff} \leq 3800$ K.}
\label{tab:2}
\end{table}

\subsection{Multiplicity}
\label{subsec:2.6}
To account for the distinct behavior of brown dwarfs in multiple systems relative to higher-mass objects, we reworked \texttt{SPISEA}'s multiplicity considerations. Surveys of nearly stellar clusters, like those done by \citet{sana2012binary} and \citet{duchene2013stellar}, have revealed that the fraction of multiple systems varies as a function of primary mass (e.g. $\sim 20\%$ for $M \sim 0.1 M_\odot$ to roughly 100\% for $M \gtrsim 5.0 M_\odot$). \texttt{SPISEA} accounts for this through the inclusion of a multiplicity class, \texttt{MultiplicityUnresolved}, that describes the multiplicity fraction (MF), companion star fraction (CSF) and mass ratio ($q$) of a simulated cluster. The MF describes the expected ratio of systems that host companions, and the CSF describes expected number of companions for a multiple system. Both are defined by \citet{reipurth1993visual} as follows:

$$\text{MF} = \frac{B + T + Q + ...}{S + B + T + Q + ...}$$

$$\text{CSF} = \frac{B + 2T + 3Q+ ...}{S + B + T + Q + ...}$$

\noindent where $S$ represents the number of single stars, $B$ the number of binary stars, $T$ the number of triple systems, $Q$ the number of quadruple systems, and so forth. The default parameters for \texttt{MultiplicityUnresolved} are given in \citet{lu2013stellar} as power-law functions of initial mass as follows:

$$\text{MF}(m) = A * m^\alpha$$
$$\text{CSF}(m) = B * m^\beta$$

\noindent where $A = 0.44$, $\alpha=0.51$, $B = 0.5$, and $\beta=0.45$, based on observations of young clusters ($< 10 \text{Myr}$).

However, observations suggest that the brown dwarf regime follows different rules. Specifically, work by \citet{aberasturi2014constraints} and \citet{fontanive2018constraining} show that multiplicity fractions decrease substantially with mass bins, and these systems are predominantly limited to binaries. With this in mind, we imposed a mass-dependent override on the \texttt{MultiplicityUnresolved} class to apply specific binary fraction values from \citet{aberasturi2014constraints}, which covers brown dwarfs of spectral types L0 to T5, and \citet{fontanive2018constraining}, which covers the remaining T5 to Y0 population. Specifically, we assert that for $0.08 \leq M/M_\odot < 0.06$, $\text{MF} = 0.16$; for $0.06 \leq M/M_\odot < 0.02$, $\text{MF} = 0.08$; and for $ M/M_\odot \leq 0.02$, $\text{MF} = 0$. The resulting MF distribution for initial masses of $0.01-10 M_\odot$ can be seen in Figure \ref{fig:10} in the Appendix. We also enforced, as per the literature, that no system with a brown dwarf primary may host more than one companion. Proof of this for a sample cluster can be seen in Figure \ref{fig:11} in the Appendix. Finally, we describe the substellar mass ratio distribution by integrating the steep power-law index $\gamma$ = 6.1 from \cite{fontanive2018constraining}, ensuring that our generated brown dwarf multiples reflect the observed preference of near-equal mass ratios. It is worth noting, however, that brown dwarfs may also be companions of higher-mass stars, which are not limited by binarity or near-equal mass ratios. 

\subsection{Binary Orbital Characteristics}
\label{subsec:2.7}

\texttt{SPISEA} contains several sub-classes to \texttt{MultiplicityUnresolved} which assigns orbital properties (such as semi-major axis, eccentricity, and inclination) to the multiple systems it generates. 
To model the orbital properties of multiple systems across the stellar and substellar regimes, we built upon the \texttt{MultiplicityResolvedDK} subclass, which is detailed in \cite{abrams2025assessing}. Specifically, we updated the mass-dependent relationship for the semimajor axis distribution based on observational constraints from \citet{duchene2013stellar} and \citet{fontanive2018constraining}, which highlight differences in the substellar regime. 

In the stellar regime ($M \gtrsim 0.08\,M_\odot$), the characteristic semimajor axis and its dispersion were described by mass-dependent analytic relations, calibrated to field binaries as per \citet{duchene2013stellar}. The mean separation was modeled using a broken power law in mass, while the logarithmic dispersion was represented as a linear function of $\log M$.

In the brown dwarf regime ($M \lesssim 0.08\,M_\odot$), observations indicate a distinct population of tighter and more narrowly distributed binaries, with typical separations of a few AU and reduced dispersion \citep{fontanive2018constraining}. To reproduce these trends, we adopted a log-normal separation distribution whose mean and standard deviation varied smoothly with mass, as per the stellar regime, but anchored the distribution in the brown dwarf regime to empirical values spanning $\sim$2–8 AU in separation and $\sigma_{\log a} \sim 0.2$–0.5.

To ensure a physically continuous transition across the hydrogen-burning limit, we interpolated between the stellar and substellar models using a smooth weighting function in $\log M$. This avoided artificial discontinuities in the semimajor axis distribution, and produced a continuous mapping of orbital properties across the full mass range. The final semimajor axis for each system is then drawn from a log-normal distribution with the mass-dependent mean and dispersion.

This approach reproduced the observed shift toward tighter, higher mass-ratio systems in the substellar regime while maintaining consistency with stellar multiplicity statistics at higher masses. Results of this interpolation can be seen in Figure \ref{fig:12} in the Appendix.

\begin{figure}
    \centering
    \includegraphics[width=\linewidth]{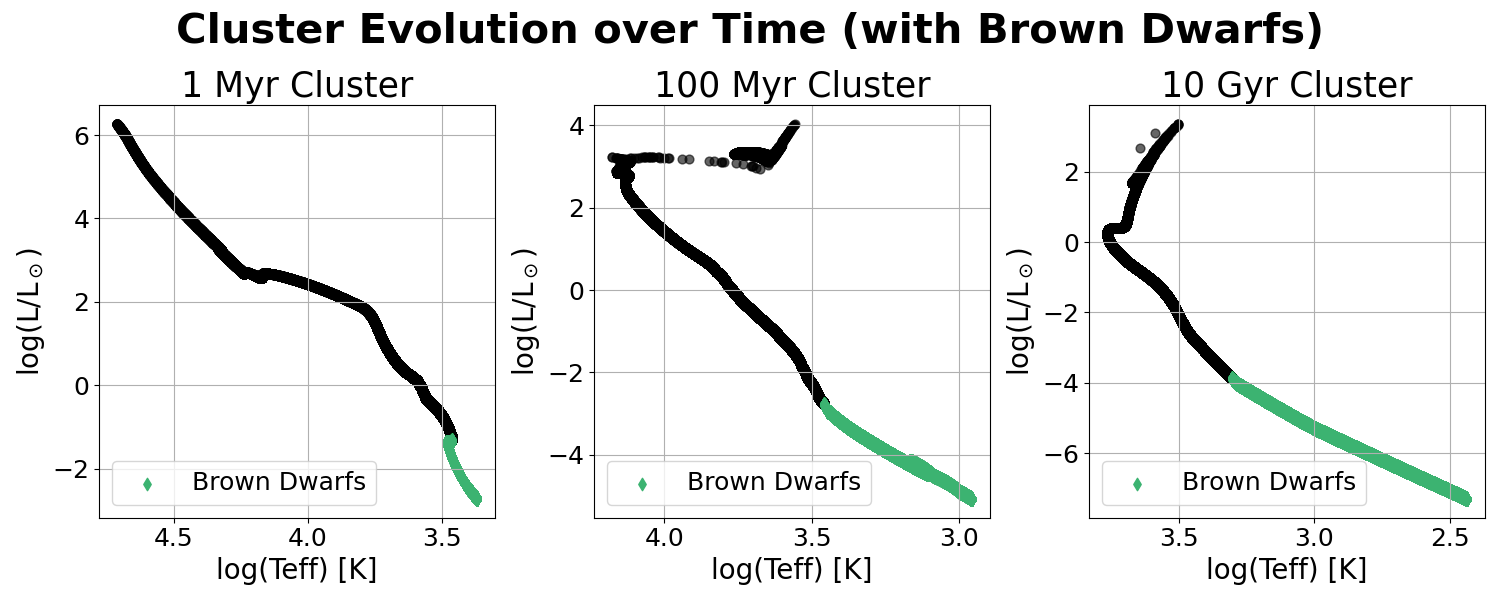}
    \caption{Hertzspring-Russell diagrams for 10$^6$ M$_\odot$ mass clusters of ages 1 Myr (left), 100 Myr (middle), 10 Gyr (right) using the newly added \texttt{MergedPhillipsBaraffePisaEkstromParsec} evolution model (Section \ref{subsec:2.5}), \texttt{Salpeter\_Kirkpatrick\_2024} initial mass function (Section \ref{subsec:2.3}), and the \texttt{get\_merged\_atmosphere} function using \cite{meisner2023exploring} atmospheric grids (Section \ref{subsec:2.4}). Brown dwarf stars are identified as having masses between 0.01 and 0.08 M$_\odot$ and are depicted with green diamonds.}
    \label{fig:5}
\end{figure}

\section{Evaluating Models Against Observed Clusters}
\label{sec:3}
After including \texttt{SPISEA} updates to model brown dwarf populations, including a new IMF, atmospheric models, evolutionary models, and multiplicity constraints, we can see how generated clusters change over time in both the stellar and substellar regimes. Figure \ref{fig:5} shows this through simulated Hertzspring-Russell diagrams of increasing ages. Expectedly, both brown dwarf and stellar populations dim over time, but brown dwarfs maintain their roughly linear shape as stellar objects deviate from the main sequence and evolve into compact remnants.

To then validate the updated brown dwarf models and confirm that the inclusion of substellar objects does not degrade the original performance of \texttt{SPISEA}, we compared simulated isochrones against well-studied open clusters: the Pleiades, Upper Sco, and M44. These comparisons follow the methodology established in \citet{hosek2020spisea}.

\subsection{Comparisons with Pleiades}
Due to its proximity and relatively low age ($\sim$ 125 Myr), the Pleiades cluster has been the subject of brown dwarf exploration efforts for decades \citep{rebolo1995discovery, moraux2003brown, pinfield2003brown}. Through telescopes like Gaia \citep{prusti2016gaia} and UKIDSS \citep{lawrence2007ukirt}, mapping of this cluster has been made possible with increasing resolution, allowing for the detection of hundreds of low-mass stars and brown dwarfs.

\begin{figure}
    \centering
    \includegraphics[width=\linewidth]{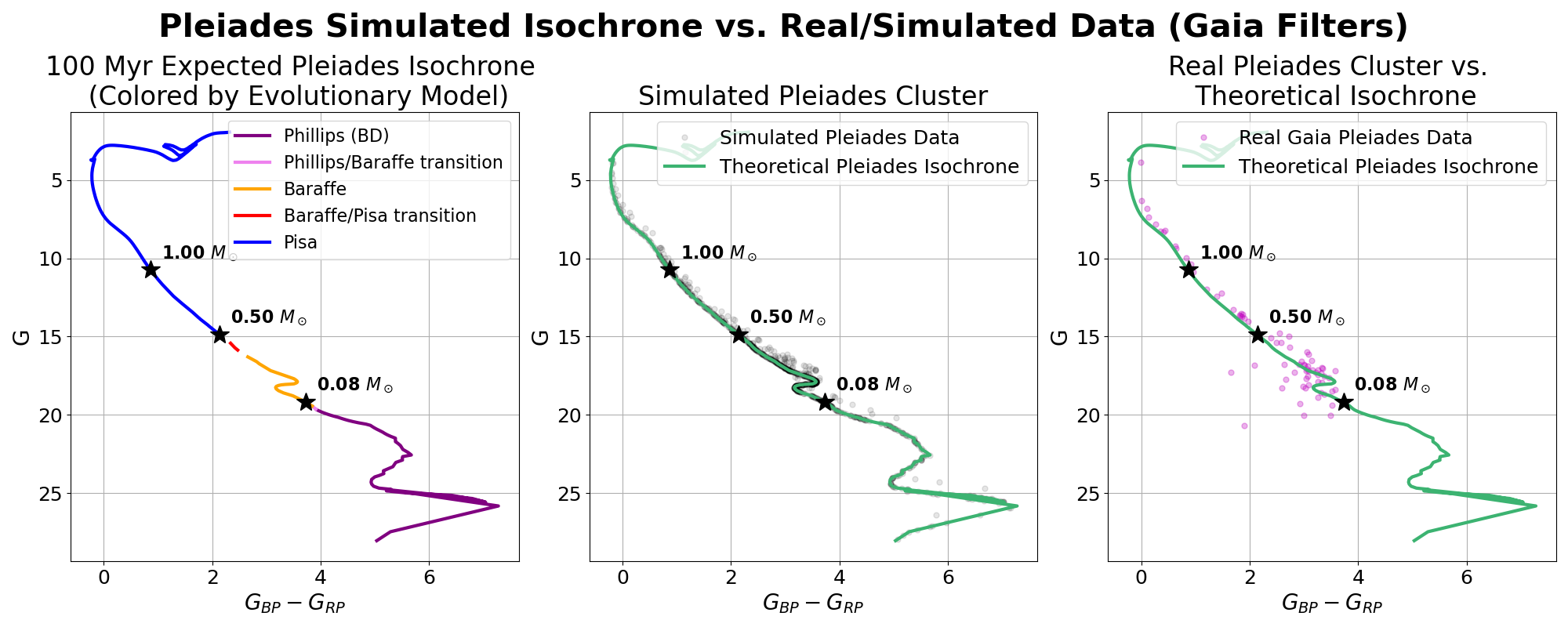}
    \caption{Comparing \texttt{SPISEA}-generated best-fit isochrones for Pleiades to simulated and real data from the Gaia mission. The left plot shows the proposed Pleiades isochrone based on cluster statistics from \cite{nagashima2003optical}, broken down by the evolutionary model (as outlined in Section \ref{subsec:2.5}). The middle plot uses the generated Pleiades isochrone to create synthetic cluster data for Pleiades, extending into the brown dwarf regime. The right plot shows the proposed isochrone against actual Pleiades data from the Gaia mission, which does not extend to the substellar-mass regime. Example values of 1.0, 0.5, and 0.08 solar masses are plotted for context.}
    \label{fig:6}
\end{figure}

Figure \ref{fig:6} presents how a simulated Pleiades cluster and isochrone generated in Gaia filters with cluster statistics (age $ \approx 125$ Myr, distance $\approx 130$ pc, reddening ($A_K$) $\approx 0.0$) from \citet{nagashima2003optical}, compares to actual Pleiades data of low and high-mass stars. Substellar objects are not highlighted in this figure, as they are fainter than the $\sim$20 magnitude threshold of the telescope \citep{prusti2016gaia}. In the Gaia $G$, $G_{\mathrm{bp}}$, and $G_{\mathrm{rp}}$ filters, the simulated isochrone closely matches the observed Pleiades sequence at higher stellar masses. This agreement demonstrates that the incorporation of substellar evolutionary models does not compromise \texttt{SPISEA}’s original performance, consistent with the results reported by \citet{hosek2020spisea}. 

\begin{figure}
    \centering
    \includegraphics[width=\linewidth]{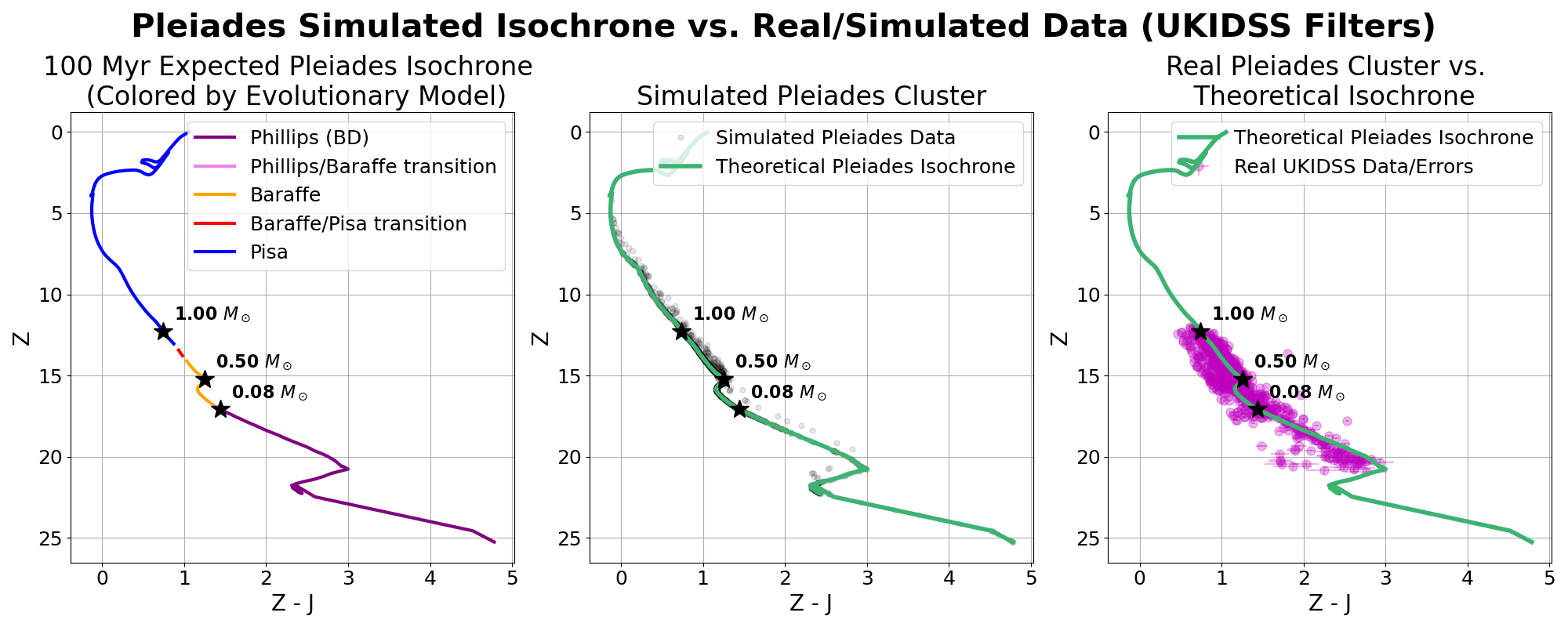}
    \caption{Comparing \texttt{SPISEA}-generated isochrones of Pleiades to observed substellar data based on \citet{lodieu2012astrometric}. Compared to Figure \ref{fig:6}, we do not observe substantial deviations in shape between the generated isochrone and observed low-mass stellar to high-mass substellar objects. However, the low-mass substellar regime shows modest deviation from observed values, indicating further exploration will be necessary as low-mass models continue to improve.}
    \label{fig:7}
\end{figure}

Figure \ref{fig:7} also presents a similarly calculated Pleiades isochrone and cluster, but instead in the $Z$ and $J$ filters of the UKIDSS mission \citep{lawrence2007ukirt}. We see that the UKIDSS $Z$ versus $Z-J$ comparison probes significantly lower masses than Gaia, including into the brown dwarf regime. The brown dwarf candidates and low-mass stars shown in the plot were provided by \citet{lodieu2012astrometric}, through photometric and proper-motion filtering. We see that the merged isochrone exhibits a smooth and continuous transition between the stellar models of \citet{baraffe2015new} and the brown dwarf models of \citet{phillips2020new}, validating our implementation processes. While the isochrone appears to preserve the shape of the observed data for moderate mass stars to high-mass brown dwarfs, modest deviations in shape are observed at the lowest masses, where photometric uncertainties increase. This suggests that brown dwarf evolutionary models are comparatively well constrained near the stellar–substellar boundary, whereas the larger deviations observed at the lowest masses are unsurprising given the scarcity of confirmed low-mass brown dwarfs and the subsequent lack of observational constraints for calibration.

\subsection{Comparisons with Upper Scorpius}
To further verify our simulated models against actual substellar data, we also considered UKIDSS photometry of the Upper Scorpius cluster. Similar to Pleiades, Upper Scorpius is young ($\sim$ 10 Myr), nearby ($\sim$ 145 pc), and work is ongoing to constrain the brown dwarf population \citep{lodieu2007new, luhman2018new, petrus2020new}. 

\begin{figure}
    \centering
    \includegraphics[width=\linewidth]{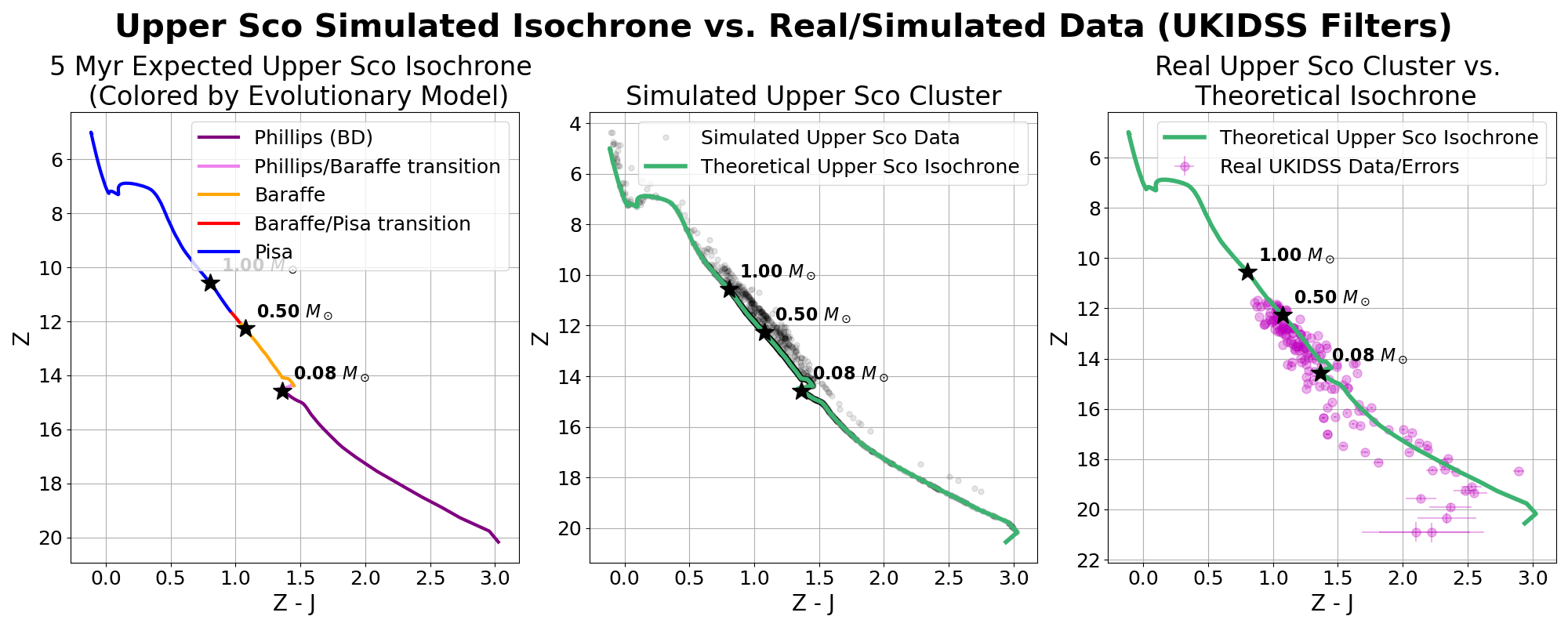}
    \caption{Comparing \texttt{SPISEA}-generated best-fit isochrones for Upper Scorpius to simulated and real data from the UKIDSS mission. The left plot shows the proposed Upper Sco isochrone based on cluster statistics from \cite{lodieu2007new}, broken down by the evolutionary model (as outlined in Section \ref{subsec:2.5}). The middle plot uses the generated Upper Sco isochrone to create synthetic cluster data for the entire mass range of Upper Sco, extending into the brown dwarf regime. The right plot shows the proposed isochrone against actual Upper Sco low-mass star and substellar data from the UKIDSS mission, provided by \citet{lodieu2007new}.}
    \label{fig:8}
\end{figure}

Figure \ref{fig:8} shows the results of the theoretical Upper Scorpius isochrone (broken down by evolutionary model), simulated cluster of all masses $\leq 7 M_\odot$, and comparison of the calculated isochrone to actual low-mass star and substellar data provided by \citet{lodieu2007new}. Similarly to Pleiades, while the models transition smoothly through the high-mass brown dwarfs, the simulated isochrone does not precisely match the curvature of the observed data for the low-mass substellar regime.

We note that, although observational errors contribute to the observed scatter in the Pleiades and Upper Scorpius clusters, they do not fully account for the systematic differences between the simulated isochrones and the UKIDSS data. These deviations are consistent with longstanding challenges in modeling young substellar objects, where uncertainties in atmospheric physics, low-gravity opacities, convection, and empirical calibration remain significant despite recent advances such as the ATMO 2020 models \citep{baraffe2002evolutionary, baraffe2003evolutionary, jeffries2012measuring, phillips2020new}. Continued observational characterization of the lowest-mass brown dwarfs will therefore be critical for improving future evolutionary and atmospheric model grids.

\subsection{Comparisons with M44}
\begin{figure}
    \centering
    \includegraphics[width=0.8\linewidth]{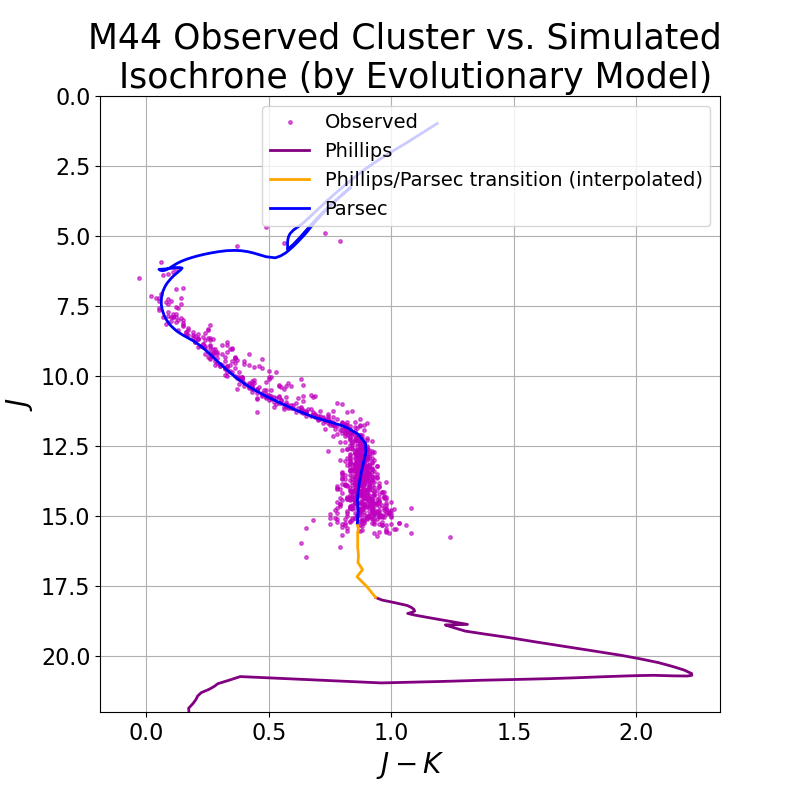}
    \caption{Comparison of the observed 2MASS CMD of the M44 cluster \citep{wang2014characterization} to a \texttt{SPISEA}-generated isochrone with best-fit values from the literature. This figure mirrors the one shown in \citet{hosek2020spisea}, proving that the original capabilities of \texttt{SPISEA} have not been compromised with the new substellar additions.}
    \label{fig:9}
\end{figure}

Finally, to validate that substellar additions have not altered the original performance of \texttt{SPISEA}, we compared the updated models to the M44 cluster, following the precedent set by \citet{hosek2020spisea}. Figure \ref{fig:9} presents the resulting $J$ versus $J-K$ color–magnitude diagram using 2MASS photometry \citep{wang2014characterization} and best-fit values from the literature. Although the observational data does not extend into the substellar regime, the simulated isochrone provides a good match across the observed mass range, and a smooth transition into lower mass regimes. This agreement demonstrates that the extended \texttt{SPISEA} framework preserves its accuracy at stellar masses while enabling physically consistent extrapolation into the brown dwarf regime.

\section{Conclusions}
\label{sec:4}
We have extended \texttt{SPISEA} to support physically consistent modeling of brown dwarfs by incorporating a modern substellar initial mass function, merged atmospheric grids, unified evolutionary tracks spanning the boundary between low-mass stars and brown dwarfs, and multiplicity and orbital behaviors. The adopted initial mass function extends smoothly into the substellar regime and reflects the observed turnover in object production near $\sim 0.05 M_\odot$, consistent with recent empirical constraints.

To enable realistic photometric predictions for brown dwarfs, we implemented merged atmospheric grids that bridge stellar and substellar temperature regimes, ensuring smooth transitions in simulated observations. In parallel, we constructed unified evolutionary tracks covering $6.0 \leq \text{log(Age/yr)} \leq 10.0$
by interpolating between established stellar models and novel substellar evolutionary grids, producing continuous relationships between mass, age, luminosity, temperature, and surface gravity across the hydrogen-burning limit.

Comparisons to observational data from the Pleiades, Upper Scorpius, and M44 clusters demonstrate that the updated framework reproduces observed stellar sequences while enabling realistic modeling of brown dwarf populations previously inaccessible within \texttt{SPISEA}. These results validate the internal consistency of the merged models and their applicability to real cluster populations. At the same time, comparisons with UKIDSS observations reveal systematic deviations between simulated, best-fit isochrones and observed photometry in the lowest-mass substellar regime. While observational uncertainties contribute to the scatter in these data, discrepancies are consistent with known limitations of current brown dwarf evolutionary and atmospheric models, which remain only weakly constrained by observations of the coolest and faintest objects. Consequently, the accuracy of simulated populations in the low-mass regime is currently limited not by the population synthesis framework itself, but by the availability and calibration of underlying theoretical models.

Collectively, these updates significantly expand the range of astrophysical problems that can be addressed using \texttt{SPISEA}, including studies of young cluster demographics and simulations of microlensing surveys. Future work should focus on both extending the theoretical model grids to lower masses and improving calibration of brown dwarf evolutionary tracks through observations of nearby young clusters and benchmark systems. Doing so will help reduce the systematic uncertainties identified in this work and enable increasingly reliable population synthesis of brown dwarfs and planetary-mass objects. As new atmospheric and evolutionary models become available, the framework developed in this work provides a basis for their incorporation into \texttt{SPISEA}, ultimately extending population synthesis studies into the planetary-mass regime.

\begin{acknowledgments}
R. Street and K. Kruszyńska gratefully acknowledge funding from the NSF grant 2206828. N.S.A. and J.R.L acknowledge support from the National Science Foundation under grant No. 1909641, the Heising-Simons Foundation under grant No. 2022-3542, and the H2H8 foundation. M.W.H. is supported by the Brinson Prize Fellowship.
\end{acknowledgments}

\begin{contribution}

C. Begbie performed the data collection, software development, modeling, and analysis, under the scientific direction of R. Street, K. Kruszyńska, and J. Lu. J. Lu, M. Hosek, and N. S. Abrams provided advice on the architecture of the \texttt{SPISEA} package and advice on implementation.


\end{contribution}

%

\software{astropy \citep{2013A&A...558A..33A, 2018AJ....156..123A, 2022ApJ...935..167A}, scikit learn \citep{scikit-learn}, \texttt{SPISEA} \citep{hosek2020spisea}, numpy \citep{harris2020array}, scipy \citep{2020SciPy-NMeth}, matplotlib \citep{Hunter:2007}, pysynphot \citep{lim2015pysynphot}}, ChatGPT \citep{chatgpt2026}


\appendix
\label{appendix}

\begin{figure}
    \centering
    \includegraphics[width=0.6\linewidth]{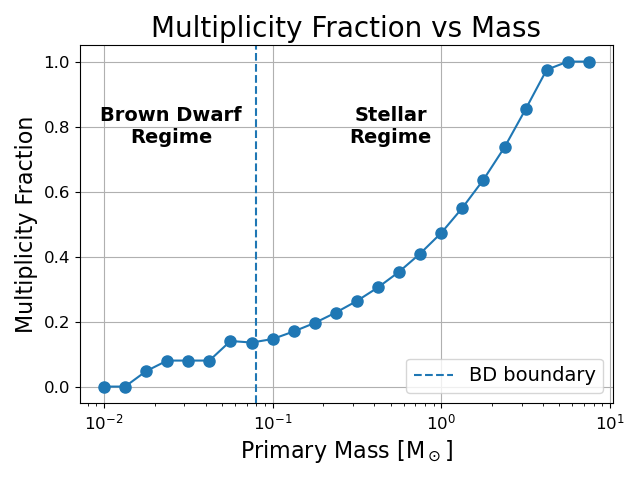}
    \caption{Multiplicity fraction from \texttt{MultiplicityUnresoved} as a function of initial mass, with the newly-implemented extension into the brown dwarf regime. The process of implementing hard-coded mass-dependent MF values for brown dwarfs is described in Section \ref{subsec:2.6} and supported by \citet{aberasturi2014constraints} and \citet{fontanive2018constraining}.}
    \label{fig:10}
\end{figure}

\begin{figure}
    \centering
    \includegraphics[width=0.6\linewidth]{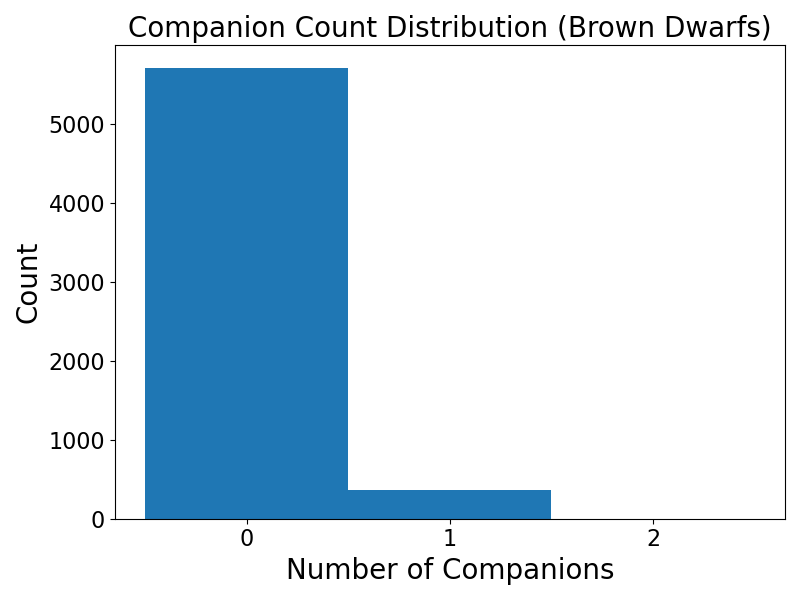}
    \caption{Companion counts for identified brown dwarf primary objects in a sample cluster. This shows the proposed relations that brown dwarf-mass objects rarely have companions, and when they do, they are limited to binaries. The process taken to implement these rules, as well as supporting works, are described in Section \ref{subsec:2.6}.}
    \label{fig:11}
\end{figure}

\begin{figure}
    \centering
    \includegraphics[width=0.6\linewidth]{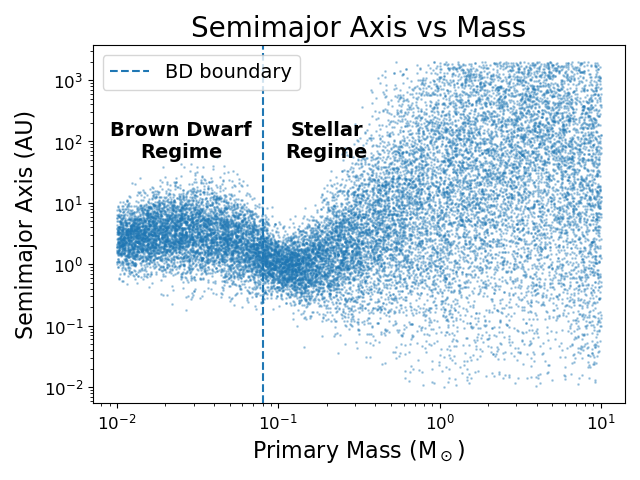}
    \caption{Extended Semimajor Axis vs. Mass distribution extending into the brown dwarf mass range, with a weighted interpolation between the substellar and stellar regimes. This is discussed in Section \ref{subsec:2.7} and helps provide orbital statistics as per the \texttt{MultiplicityResolvedDK} class.}
    \label{fig:12}
\end{figure}

\bibliography{sample7}{}
\bibliographystyle{aasjournalv7}



\end{document}